\documentclass[conference]{IEEEtran}

\usepackage[hyphens]{url}
\usepackage{hyperref}
\usepackage{mathtools}
\usepackage{amsmath}
\usepackage{orcidlink}
\hypersetup{pdfborder={0 0 0}}

\usepackage{booktabs, tabularx}
\usepackage[referable]{threeparttablex}
\usepackage{stfloats}
\usepackage{enumitem}
\usepackage{cite}
\usepackage{amssymb,amsfonts}
\usepackage{mathabx}
\usepackage{algorithmic}
\usepackage{graphicx}
\usepackage{textcomp}
\usepackage{xcolor}

\usepackage{array,ragged2e}
\newcolumntype{R}[1]{>{\RaggedLeft\arraybackslash}p{#1}}

\def\BibTeX{{\rm B\kern-.05em{\sc i\kern-.025em b}\kern-.08em
    T\kern-.1667em\lower.7ex\hbox{E}\kern-.125emX}}
    
\begin{document}
\title{Vision: An Extensible Methodology for Formal Software Verification in Microservice Systems
}
\author{\IEEEauthorblockN{1\textsuperscript{st} Connor Wojtak 
\orcidlink{0009-0007-9094-138X}}
\IEEEauthorblockA{\textit{Electrical and Computer Engineering} \\
\textit{University of Arizona}\\
Tucson, Arizona, USA \\
connorwojtak@arizona.edu}
\and
\IEEEauthorblockN{2\textsuperscript{nd} Darek Gajewski 
\orcidlink{0009-0005-0441-4213}}
\IEEEauthorblockA{\textit{Electrical and Computer Engineering} \\
\textit{University of Arizona}\\
Tucson, Arizona, USA \\
dgajewski@arizona.edu}
\and
\IEEEauthorblockN{3\textsuperscript{rd} Tomas Cerny 
\orcidlink{0000-0002-5882-5502}}
\IEEEauthorblockA{\textit{Electrical and Computer Engineering} \\
\textit{University of Arizona}\\
Tucson, Arizona, USA \\
tcerny@arizona.edu}
}

\maketitle

\renewcommand{\thefootnote}{}
\footnotetext{
© 2025 IEEE. Personal use of this material is permitted. Permission from IEEE must be obtained for all other uses, in any current or future media, including reprinting/republishing this material for advertising or promotional purposes, creating new collective works, for resale or redistribution to servers or lists, or reuse of any copyrighted component of this work in other works.
}
\renewcommand{\thefootnote}{\arabic{footnote}}

\begin{abstract}
Microservice systems are becoming increasingly adopted due to their scalability, decentralized development, and support for continuous integration and delivery (CI/CD). However, this decentralized development by separate teams and continuous evolution can introduce miscommunication and incompatible implementations, undermining system maintainability and reliability across aspects from security policy to system architecture. We propose a novel methodology that statically reconstructs microservice source code into a formal system model. From this model, a Satisfiability Modulo Theories (SMT) constraint set can be derived, enabling formal verification. Our methodology is extensible, supporting software verification across multiple cross-cutting concerns. We focus on applying the methodology to verify the system architecture concern, presenting formal reasoning to validate the methodology's correctness and applicability for this concern. Additional concerns such as security policy implementation are considered. Future directions are established to extend and evaluate the methodology.
\end{abstract}

\begin{IEEEkeywords}
Microservices, Verification, Static Analysis, System Modeling, SMT Solving, Maintainability, Decentralization
\end{IEEEkeywords}
\vspace{-1em}

\section{Introduction}

The integration of cloud-native systems within modern enterprises has redefined the landscape of software architecture, particularly through the adoption of Microservice Architecture (MSA)~\cite{cerny2018microservice}. However, while considerable work has been done to model and formally verify monolithic and cloud-native systems such as in reviews by Jawaddi et al.~\cite{jawaddi2022formalReview} and Souri et al.~\cite{souri2018formal}, we find little work that performs formal software verification in microservices systems at the application layer. 


Microservices, unlike monolithic and cloud-native systems, are generally developed by separate development teams in separate repositories \cite{amoroso2023msdev}. This fragmented approach to development means that developers lack a complete overview of the system, which increases the burden for cross-cutting concerns such as security, business, and system architecture. This is compounded by a lack of centralized governance in microservice systems to verify concerns are properly addressed \cite{cerny2018microservice, zhou2023microservicepain}. While solutions such as end-to-end (E2E) testing or static analysis exist for addressing concerns like verifying security policies or analyzing system architecture, these existing approaches have significant limitations and costs \cite{smith2023endtoendbenchmark}.

Considering E2E tests, microservice systems inherently use more endpoints and connections due to their basis on interservice communication. Thus, a more extensive suite of E2E tests is needed to comprehensively verify the entire system. This places significant burden on development and operations teams to maintain full coverage for E2E service verification. Microservices also often undergo continuous small, iterative changes \cite{cerny2018microservice}, meaning tests can easily fall out of date or fail to provide complete coverage, making complete verification even more time consuming for developers. In addition, comprehensive testing can be compute-intensive, which can incur significant cost as microservice systems scale \cite{smith2023endtoendbenchmark}. 

While some existing static analysis techniques address these shortcomings, they also have limitations. They may focus on one concern and check rules exclusively rather than simultaneously to ensure a system satisfies a concern \cite{cerny2025changeimpactanalysismicroservicesbased}.


We propose a static and extensible methodology for formal software verification in microservice systems by reverse engineering system source code into an intermediate formal model. From this, an extensible constraint model can be generated that unifies rules for cross-cutting concerns. Using the constraint model, an SMT solver can verify a system satisfies all constraints for every concern considered or modify the system to satisfy them. This approach addresses limitations in both existing static analysis and testing methods, revolutionizing application-level software verification in microservice systems by reducing developer burden, development costs, and testing costs through the application of formal methods.





This paper is organized as follows. Section \ref{sec:Background} elaborates on the problem’s background and related work. Section \ref{sec:Methodology} identifies our methodology. Section \ref{sec:Proofs} presents formal discussion and validation of our methodology. Section \ref{sec:research-questions} outlines our future research questions. Section \ref{sec:Conclusion} concludes our paper.   

\section{Background and Related Work}\label{sec:Background}

With the widespread adoption of MSA, tools to perform dynamic analysis and static analysis for MSA have been growing in number. Analysis or testing of software systems serves multiple aspects, mainly to keep a system reliable and available. The DevOps Research and Assessment (DORA) metrics, which measure the frequency of deployments and issues, are critical for any enterprise to measure. These four metrics are the frequency of deployments, the number of issues caused by deployments, the lead time needed for a successful push to production, and the mean time to recovery \cite{daraojimba2024DevOpsMetrics}.

Dynamic analysis, which involves real-time testing or profiling of systems, has been used to detect issues or architectural degradation in the system. 
Maruf et al. \cite{Maruf2022TelemtryDynamic} combine dynamic analysis tools with anomaly detection, utilizing quality metrics and anti-patterns, to detect architectural degradation in microservice systems. 
Sotomayor et al. \cite{Sotomayor2019RuntimeTesting} make use of runtime testing in microservices and look at different levels of testing tools for unit testing, performance testing, and system verification. However, this can be expensive, especially with MSA, as the log volume and telemetry data grows with the number of services within the microservice environment \cite{smith2023endtoendbenchmark}.  

While static analysis methods cannot replace specific testing, such as performance testing of systems, the use of static analysis can be employed to perform software verification at a significantly lower cost and time, leading to the discovery of issues before they are deployed into real-time environments. 

Existing tools, such as CIMET \cite{cerny2025changeimpactanalysismicroservicesbased}, can utilize static analysis to identify architecture rules that are violated across multiple microservices repositories. However, a tool like this can only consider antipatterns one at a time when scanning the code repository, and is also unable to suggest changes. Our tool will be able to consider all the rules simultaneously, allowing for fixes to incorporate these rules without introducing further degradation in the system. Cerny et al. \cite{cerny2023catalog} catalog antipatterns in microservices, which tools like CIMET \cite{cerny2025changeimpactanalysismicroservicesbased} can detect. Abdelfattah et al. \cite{abdelfattah2023towards} apply ReSSa, a predecessor to CIMET for static analysis, to enumerate data entity accesses along endpoint call chains to calculate access rights.

Formal methods have been previously used for software verification. However, methods such as Armando et al. \cite{armando2010automatedsymbolicanalysisarbacpolicies}, which formalizes methods for authorization policy consistency, or Lattuada et al. \cite{Lattuada2024FoundationForSysVerification}, which makes use of a solver for eliminating bugs in software, are designed to address issues in traditional monolithic systems.

There are some cloud-native solutions such as how Backes et al. \cite{Backes2018FormalIAM} utilizes AWS' Zelkova system and formal methods to verify IAM policies. Additionally, Panda et al. \cite{panda2017SVinmicroservices} discusses formal methods for verifying communication behaviors between microservices. However, these solutions are at the infrastructure layer and network layers, respectively --- not at the application layer working with the system source code. This is echoed by other systematic literature reviews such as Souri et al. \cite{souri2018formal} or Jawaddi et al. \cite{jawaddi2022formalReview} that identify little to no work on formal software verification in microservice systems at the application layer. Our methodology aims to fill this gap.

\section{Methodology}\label{sec:Methodology}
\begin{figure*}
    \centering
    \includegraphics[width=1\linewidth]{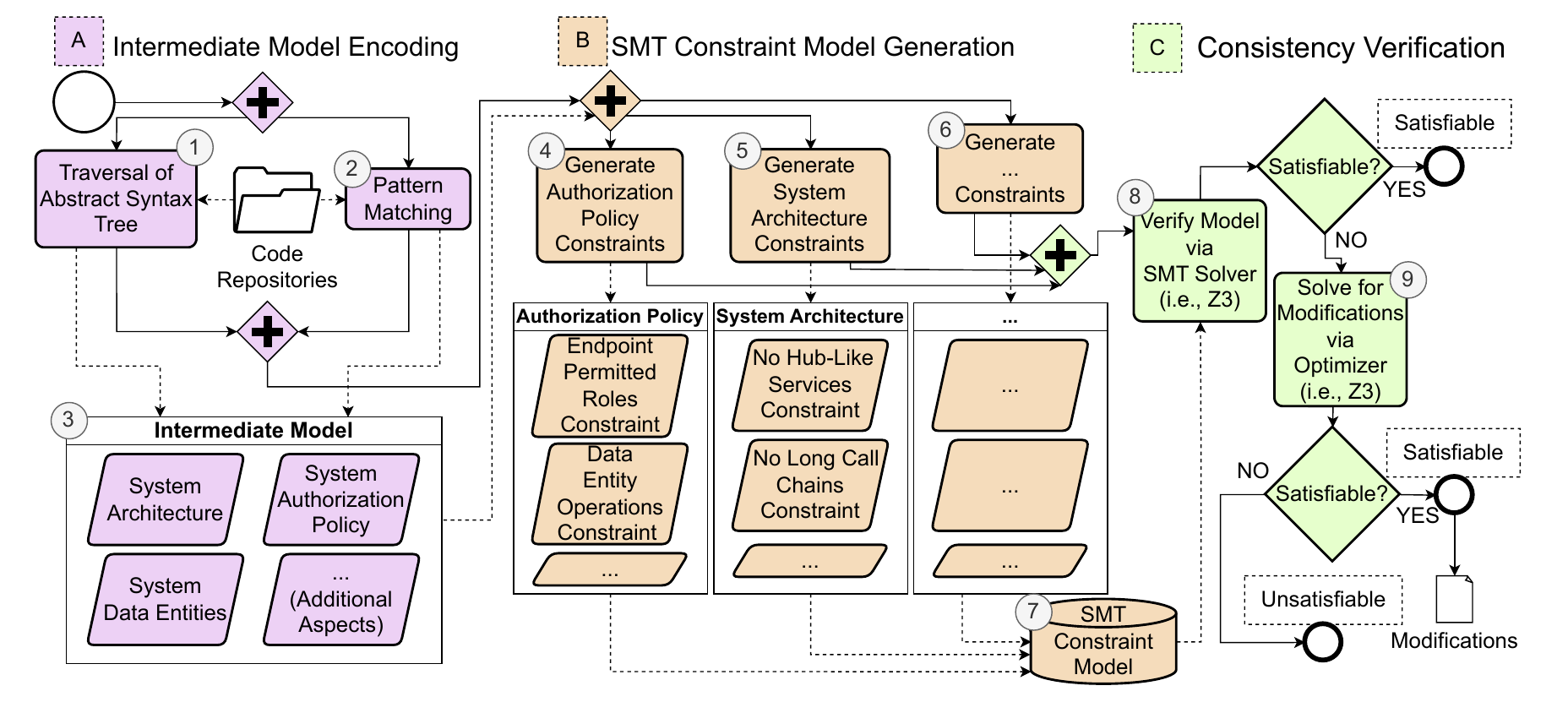}
    \vspace{-2.25em}
    \caption{An overview of the verification process, starting with the creation of an intermediate model from which various constraint sets can be generated. These sets form a constraint model that can be supplied to a solver to verify the microservice system.}
    \label{fig:overall-process}
    \vspace{-1em}
\end{figure*}

Formally verifying a microservice system requires encoding aspects of the system into an SMT constraint model that can be supplied to any SMT solver such as \texttt{Z3} or \texttt{CVC5}. We propose a three-stage process, as shown in Figure~\ref{fig:overall-process}, in which we create an intermediate model, create an SMT constraint model, and perform consistency verification. The intermediate model encodes various aspects of the system such as architecture, authorization policy, data entities, and other additional aspects. These are then traversed to generate constraints to enforce rules for various concerns, which are supplied to an SMT solver to determine if the system satisfies the generated rules. We focus on describing encodings and constraints for verifying the system architecture concern shown in Figure~\ref{fig:overall-process}. We also discuss possible extensions for the authorization policy concern. We describe the encoding of the intermediate model, generation of the SMT constraint model, and constraint solving in Sections~\ref{sec:intermediate-model-encoding}, \ref{sec:constraint-model-gen}, and \ref{sec:consistency-verification}, respectively.

\subsection{Intermediate Model Encoding}\label{sec:intermediate-model-encoding}
\subsubsection{Model Definition}\label{sec:model-def}
Applying static analysis techniques on the microservice system code repositories, such as traversing an abstract syntax tree or performing pattern matching, allows for the extraction of the system architecture, data entities, and security policies ((1) and (2) in Figure~\ref{fig:overall-process}). We then create an intermediate model ((3) in Figure~\ref{fig:overall-process}) where we formally define a microservice system as a tuple with four core attributes, extensible by an arbitrary number of additional attributes:
\[
\mathcal{S} = (\mathcal{M}, \mathcal{E}, \texttt{E\_parent}, \mathcal{G}, \ldots)
\]
Where:

{
\begin{raggedright}
\begin{description}
    \item $\mathcal{M} = \{MS_1, \ldots, MS_n\}$ is the set of microservices.
    \item $\mathcal{E} = \{E_1, \ldots, E_m\}$ is the set of all endpoints. Many endpoints belong to a single microservice.
    \item $\texttt{E\_parent} : \mathcal{E} \rightarrow \mathcal{M}$ maps an endpoint to its single parent microservice.
    \item $\mathcal{G} \subseteq \mathcal{E} \times \mathcal{E}$ is the directed graph representing remote service calls between endpoints.
\end{description}
\end{raggedright}
}

Now consider the concern that a microservice system satisfies a set of architectural rules (ARs) to prevent antipatterns from forming. Examples of such antipatterns could include cyclic dependencies between endpoints or microservices with a large number of inbound or outbound connections. These rules primarily depend on the endpoints and service calls between them; however, an extension of the model is necessary to identify cyclic dependencies. Thus, the following extension of $\mathcal{S}$, $\mathcal{S}_{AR}$, can be defined as follows:
\begin{align*}
\mathcal{S}_{AR} = (\mathcal{M}, \mathcal{E}, \texttt{E\_parent}, \mathcal{G}, \mathcal{L}, \texttt{L\_endpoint})
\end{align*}
Where:
\begin{raggedright}
\begin{description}
    \item $\mathcal{M}$, $\mathcal{E}$, $\texttt{E\_parent}$, and $\mathcal{G}$ are the same as defined in $\mathcal{S}$.
    \item $\mathcal{L} = (\ell_1,\ldots\ell_m)$ defines the topological ordering of endpoints where $m$ is the size of $\mathcal{E}$ and $\forall i \in \{1,\ldots,m\}$, $\ell_i \in \mathcal{Z}$ and $\mathcal{Z}$ is the set of integers.
    \item $\texttt{L\_endpoint} : \mathcal{E} \rightarrow \mathcal{L}$ maps an endpoint $E_i$ to its ordering $L_i$.
\end{description}
\end{raggedright}

On the other hand, consider the concern of enforcing authorization policy at the endpoint and data entity levels in a microservice system. Then, the following extension of $\mathcal{S}$, $\mathcal{S}_{authorization}$, can be defined as follows:
\begin{align*}
\mathcal{S}_{authorization} = (\mathcal{M}, \mathcal{E}, \texttt{E\_parent}, \mathcal{G}, \mathcal{D}, \mathcal{A}, \mathcal{O}, \\\texttt{A\_operations}, \mathcal{R}, \texttt{E\_permitted}, \ldots)
\end{align*}
Where:
{
\begin{raggedright}
\begin{description}
    \item $\mathcal{M}$, $\mathcal{E}$, $\texttt{E\_parent}$, and $\mathcal{G}$ are the same as defined in $\mathcal{S}$.
    \item $\mathcal{D} = \{D_1, \ldots, D_h\}$ is the set of data entities.
    \item $\mathcal{A} = \{A_1, \ldots, A_j\}$ is the set of data entity accesses for an entity in $\mathcal{D}$ (where an entity is created, read, etc.). There is at most one data entity access for each data entity in $\mathcal{D}$ for each endpoint in $\mathcal{E}$.
    \item $\mathcal{O} = \{O_1, \ldots, O_k\}$ is the set of possible operations that can be performed on a data entity (often consisting of four entries: create, read, update, and delete).
    \item $\texttt{A\_operations} : \mathcal{A} \rightarrow \mathcal{P}(\mathcal{O})$ maps an instance of a data entity access to the set of operations the access performs (where $\mathcal{P}(\mathcal{O})$ is the power set of $\mathcal{O}$).
    \item $\mathcal{R} = \{R_1, \ldots, R_\ell\}$ is the set of possible roles users can assume in the microservice system.
    \item $\texttt{E\_permitted} : \mathcal{E} \rightarrow \mathcal{P}(\mathcal{R})$ maps an endpoint to the set of roles (where $\mathcal{P}(\mathcal{R})$ is the power set of $\mathcal{R}$).
\end{description}
\end{raggedright}
}

These two extensions can be combined into a single extensive model $\mathcal{S}_{combined}$ for authorization policy and architectural rule verification. This could be defined as:
\begin{align*}
\mathcal{S}_{combined} = (\mathcal{M}, \mathcal{E}, \texttt{E\_parent}, \mathcal{G}, \mathcal{L}, \texttt{L\_endpoint}, \mathcal{D}, \\\mathcal{A}, \mathcal{O}, \texttt{A\_operations}, \mathcal{R}, \texttt{E\_permitted}, \ldots)
\end{align*}
where each variable listed is as defined previously in $\mathcal{S}$, $\mathcal{S}_{AR}$, and $\mathcal{S}_{authorization}$.

As needed, additional variables could be added to address further concerns. For example, variables such as fields or attributes of data entities can be added to the model to enable the enforcement of business rules related to personally identifiable information in later steps. This model makes some simplifications compared to a real-world microservice. For example, $\mathcal{G}$ is simplified in that endpoints might make remote service calls based on a condition rather than on every invocation of an endpoint. However, for our constraints addressing the authorization policy and architecture concerns, whether a service call is conditional is irrelevant. Instead, the potential for a service call to be made from one endpoint to another is enough to identify it as a call in $\mathcal{G}$. However, if a constraint were to arise requiring information on conditional service calls, $\mathcal{S}$ could be extended to include which calls are made only under particular conditions. Thus, simplifications made by the model can likely be addressed by extending the model.

\subsubsection{Model Representation}\label{sec:model-rep}
Having defined the formal model $\mathcal{S}$ and its extensions, we now describe how this model is encoded into a solver-compatible format so that an SMT constraint model can then be generated.

For $\mathcal{S}$:
{
\begin{raggedright}
\begin{description}
    \item $\mathcal{M}$ can be represented as a set $M$ where $M_i \in M$ is a subset of distinct endpoints from $\mathcal{E}$ where $i \in [0,n-1]$.
    \item $\mathcal{E}$ can be represented as a set $E$ where each $E_i \in E = i$ where $i$ is an integer and $i \in [0,m-1]$.
    \item $\texttt{E\_parent}$ can be represented as an array of integers $E_{parents}$ where each element $E_{parents_i}$ is an integer $j \in [0,n-1]$ such that $E_{parents}(i) = j$, where $i \in [0,m-1]$ corresponds to an endpoint in $\mathcal{E}$ and $j$ corresponds to the endpoint's parent microservice in $\mathcal{M}$.
    \item $\mathcal{G}$ can be represented as an adjacency list or matrix $G$ such that $G[i][j] = True$ iff there is an edge from $E_i$ to $E_j$ for $i \in \mathcal{E}, j \in \mathcal{E}$. Otherwise $G[i][j] = False$.
\end{description}
\end{raggedright}
}

For $\mathcal{S}_{AR}$:
{
\begin{raggedright}
\begin{description}
    \item $\mathcal{L}$ can be represented as an integer array $L$ of size $m$.
    \item $\texttt{L\_endpoint}$, in practice, is subsumed by the array $L$ where the topological order of endpoint $E_i$, is simply given by $L[i]$.
\end{description}
\end{raggedright}
}


\subsection{SMT Constraint Model Generation}\label{sec:constraint-model-gen}
Once the microservice system is encoded into an intermediate model ($\mathcal{S}$ or one of its extended versions), the intermediate model can be traversed to generate constraints that address various concerns (such as (4), (5), and (6) in Figure~\ref{fig:overall-process}). 

We provide examples of constraints that can be enforced for the system architecture concern, which focuses on preventing common microservice antipatterns \cite{cerny2023catalog}. From this catalog of antipatterns, we formally define three constraints to identify and prevent hub-like dependencies and cyclic dependencies. For $G$, we consider only the edges that are present in the system at the time of analysis. We do not allow the solver to suggest adding edges to the system (only removing existing ones), as traversing the entire $\mathcal{E} \times \mathcal{E}$ graph search space would be intractable. 

\subsubsection{No Intraservice Remote Calls Constraint}
In a microservice system, generally no endpoint should make a remote service call to another endpoint in the same microservice; instead, they should invoke service functions programmatically. This also subsumes a rule that no endpoint can make a remote service call to itself. The constraint enforcing this rule, $C_{nirc}$, is defined as follows based on the model $\mathcal{S}_{combined}$:
\begin{align*}
C_{nirc} = \bigwedge_{(e_1, e_2) \in G} \texttt{E\_parent}(e_1) \ne \texttt{E\_parent}(e_2)
\end{align*}

\subsubsection{No Hub-like Dependencies Constraint}
In a microservice system, a hub-like dependency introduces centralization, which can lead to scalability issues in a microservice system and introduce a single point of failure as many services depend on it \cite{cerny2023catalog}. This translates into our problem space as a microservice having a large number of endpoints with many inbound and outbound connections. The constraint preventing these hub-like dependencies, $C_{hub}$, is defined as follows based on the model $\mathcal{S}_{combined}$:
\begin{align*}
C_{hub} = \bigwedge_{M_i \in M}[\sum_{E_j \in M_i, e \in E}(conn_{E_j,e} + conn_{e,E_j})] > \tau
\end{align*}
Where:
\begin{raggedright}
\begin{description}
\item $conn_{E_a,E_b} = 1$ iff $G[a][b] = True$ and $conn_{E_a,E_b} = 0$ otherwise.
\item $\tau$ is a user-defined threshold that, for higher values, permits more dependency on a single microservice, and for lower values, permits less dependency on a single microservice.
\end{description}
\end{raggedright}

\subsubsection{No Cyclic Dependencies Constraint}
In a microservice system, a cyclic dependency occurs when a call to an endpoint triggers a chain of remote service calls that ultimately lead back to the original endpoint \cite{cerny2023catalog}. More formally, it means that $G$ contains a cycle and cannot be topologically sorted. The constraint preventing cyclic dependencies, $C_{cycle}$, is defined as follows based on the model $\mathcal{S}_{combined}$:
\begin{align*}
C_{cycle} = \bigwedge_{(e, e') \in G} L[e] < L[e']
\end{align*}

\subsubsection{Constraints for Other Concerns}

While we only formally define constraints for the architectural concern, constraints can be generated for other concerns. For example, consider the concern regarding authorization policies. Constraints could be generated to ensure roles have consistent access to data entities across microservices and constraints to ensure authorized roles can complete call chains. Using the extended model $\mathcal{S}_{authorization}$, it is feasible to enforce said constraint, as it contains information about  permitted roles to endpoints, data entity operations, and relationships between all the endpoints, data entities, etc.


As the constraints are generated, they can be stored together to form a unified SMT constraint set ((7) in Figure~\ref{fig:overall-process}). When generating the architectural and security constraints, we hold all sets of variables in the model constant except for the system connections ($\mathcal{G}$) and the endpoint permitted roles (\texttt{E\_permitted}). Variables are held constant either by directly creating a constraint that ensures the variable matches the actual system or by generating constraints programmatically on the assumption that a variable is held constant. When considering different concerns, different combinations of variables could be held constant based on which aspects of the system should be changed to address the concern.

\subsection{Consistency Verification}\label{sec:consistency-verification}
After the constraint set is complete, an SMT solver ((8) in Figure~\ref{fig:overall-process}) can be used to verify that the microservice system meets the defined rules or constraints. We can create a verifier by taking a constraint system and holding all variables constant via constraints that assert the variables match the actual state of the microservice system. This verifier can determine whether the existing system satisfies the constraints for each concern.

On the other hand, if the existing system fails to satisfy the constraints for each concern, an optimizer ((9) in Figure ~\ref{fig:overall-process}) such as Z3's optimizer can take the original constraint system and solve for a set of variables that makes the system conform to the constraints, with an objective function that minimizes the number of changes that made to the existing microservice system. This objective function was selected as, while it may suggest changes that are not optimal due to missing high-level context or reasoning (i.e., suggesting regular users be able to access an admin-only endpoint to satisfy a constraint), it provides locations in the system for developers to investigate the best solution to the inconsistency.

As the system scales to have a large number of microservices and endpoints, or the number of concerns simultaneously considered increases, the number of constraints generated could feasibly become intractible to satisfy. In these situations, it may be prudent to consider the constraint set for each concern separately to maintain efficient running time. While this limits the solver's ability to identify a set of variables satisfying constraints across all concerns, for large microservice systems, the methodology remains beneficial for identifying and resolving issues for each separate concern.


\section{Formal Reasoning and Soundness}\label{sec:Proofs}
We present a formal discussion of the base intermediate model $S$, the intermediate model $S_{AR}$, and the SMT constraint model for addressing the architectural concern.

\subsection{Intermediate Model Soundness}\label{sec:Soundness}
\subsubsection{Extracting Model Information}
\begin{table}[t!]
\scriptsize
    \caption{Model to System Traceability}
    \begin{center}
    \begin{tabular}{|p{0.4\linewidth}|p{0.4\linewidth}|}
    \hline
    \textbf{Model Component}&\textbf{Extracted From}\\
    \hline
    $\mathcal{M}$: Microservices&Project structure, manifest files, build files (i.e., a pom.xml file)\\
    \hline
    $\mathcal{E}$: Endpoints&REST controller definitions and route annotations/decorators\\
    \hline
    $\texttt{E\_parent}$: Endpoint Parents&Location of controller definitions in project structure\\
    \hline
    $\mathcal{G}$: Call Graph Edges&Traversal of abstract syntax tree connecting endpoints and REST calls (with REST calls identified via framework-specific calls)\\
    \hline
    $\mathcal{L}$: Ordering of Endpoints&Not extracted from the system --- used by the SMT solver to perform the topological ordering\\
    \hline
    $\texttt{L\_endpoint}$: Endpoint Ordering&Not extracted from the system --- used by the SMT solver to map an endpoint to its ordering\\
    \hline
    \end{tabular}
    \label{tab:model_traceability}
    \end{center}
    \vspace{-3em}
\end{table}

As a first step in validating our intermediate model, we identify how the information in the model can be extracted from the source code. Table~\ref{tab:model_traceability} illustrates each component of $\mathcal{S}_{AR}$ (and thus also $\mathcal{S}$) and where each component can be extracted from in the source code. This does highlight some potential shortcomings. Namely, the model will only be as accurate as the static analysis that extracts information from the source code. While static analysis can prepare a mostly accurate representation, it is difficult, if not impossible, to generate an exact call graph. Furthermore, the static analysis tool generally needs to target a specific programming language, framework, and/or libraries. 

However, based on Table~\ref{tab:model_traceability}, we can see that it is possible to extract all of the necessary information to a reasonable degree of accuracy. Furthermore, the extraction of the information is generalizable to any language with an abstract syntax tree, meaning that obtaining a highly accurate intermediate model of the system is a matter of developing an accurate enough tool to statically extract the necessary information for each programming language, framework, etc. Once a tool is developed for a specific language or framework, it could be used by all projects developed with that language or framework. Therefore, we conclude that the definition of our model is based on real microservice system constructs and that we can extract the information necessary to construct the model.

\subsubsection{Encoding Model Information}

To further support the soundness of our intermediate model, we manually work out how a representative subset of the TrainTicket microservice system benchmark can be encoded into our formal model. We selected TrainTicket \cite{train-ticket-2018} as our benchmark from a dataset of microservice system benchmarks \cite{amoroso2024dataset} due to its size, the number of system connections, and its focus on representing a real-world microservice system \cite{train-ticket-2018} \footnote{The TrainTicket repository used is available at \url{https://github.com/FudanSELab/train-ticket}. We use the code from commit ID \texttt{313886e99befb94be6cd45f085c98e0019f59829}.}. We select three of 45 microservices to encode into our model --- \texttt{ts-admin-basic-info-service}, \texttt{ts-contacts-service}, and \texttt{ts-price-service}. We chose these services because they have numerous system connections and are representative of the rest of the system.

We classify \texttt{ts-admin-basic-info-service} as $M_0$ and identify 21 endpoints; therefore, $M_0 = \{E_0, \ldots, E_{20}\}$. Meanwhile, $E_{parents}[i] = 0, \forall i \in [0, 20]$, with $i$ corresponding to $E_i$ and $E_{parents}[i]$ corresponding to $M_0$.

We classify \texttt{ts-contacts-service} as $M_1$ and identify 8 endpoints; therefore $M_1 = \{E_{21}, \ldots, E_{28}\}$. Meanwhile, $E_{parents}[i] = 1, \forall i \in [21, 28]$, with $i$ corresponding to $E_i$ and $E_{parents}[i]$ corresponding to $M_1$.

We classify \texttt{ts-price-service} as $M_2$ and identify 7 endpoints; therefore $M_2 = \{E_{29}, \ldots, E_{35}\}$. Meanwhile, $E_{parents}[i] = 2, \forall i \in [29, 35]$, with $i$ corresponding to $E_i$ and $E_{parents}[i]$ corresponding to $M_2$.

Since this is the subset of microservices we are considering, $M = \{M_0, M_1, M_2\}$. By definition, each $E_i = i$; therefore, $E = \{0,1,\ldots,35\}$. Between these endpoints, we identify multiple connections from \texttt{ts-admin-basic- info-service} to \texttt{ts-contacts-service} and \texttt{ts- price-service}, which are shown in Table~\ref{tab:system_connections}. Based on these connections, for every $E_i \Rightarrow E_j$ in Table~\ref{tab:system_connections}, $G[i][j]=True$. This completes the construction of $\mathcal{S}$ from TrainTicket's source code.

Since $L$ is not extracted from the microservice system but is instead a transitive element used only at the solver level to perform topological ordering of endpoints, $L$ can be initialized to an array of all zeroes of length $m$, one entry for each endpoint. This completes the construction of $\mathcal{S}_{AR}$ from TrainTicket's source code.

\begin{table}[t!]
\scriptsize
    \caption{Identified System Connections}
    \vspace{-2em}
    \begin{center}
    \begin{tabular}{|>{\raggedright\arraybackslash}p{0.74\linewidth}|>{\centering\arraybackslash}p{0.16\linewidth}|}
    \hline
    \textbf{Endpoint Paths}&\textbf{Encoding}\\
    \hline
    \texttt{GET /adminbasic/contacts} to \texttt{GET /contactservice/contacts}&$E_0 \Rightarrow E_{21}$\\
    \hline
    \texttt{DELETE /adminbasic/contacts/\{contactsId\}} to \texttt{DELETE /contactservice/contacts/\{contactsId\}}&$E_1 \Rightarrow E_{22}$\\
    \hline
    \texttt{PUT /adminbasic/contacts} to \texttt{PUT /contactservice/contacts}&$E_2 \Rightarrow E_{23}$\\
    \hline
    \texttt{POST /adminbasic/contacts} to \texttt{POST /contactservice/contacts/admin}&$E_3 \Rightarrow E_{24}$\\
    \hline
    \texttt{GET /adminbasic/prices} to \texttt{GET /priceservice/prices}&$E_4 \Rightarrow E_{29}$\\
    \hline
    \texttt{DELETE /adminbasic/prices/\{pricesId\}} to \texttt{DELETE /priceservice/prices/\{pricesId\}}&$E_5 \Rightarrow E_{30}$\\
    \hline
    \texttt{PUT /adminbasic/prices} to \texttt{PUT /priceservice/prices}&$E_6 \Rightarrow E_{31}$\\
    \hline
    \texttt{POST /adminbasic/prices} to \texttt{POST /priceervice/prices/admin}&$E_7 \Rightarrow E_{32}$\\
    \hline
    \end{tabular}
    \label{tab:system_connections}
    \end{center}
    \vspace{-3em}
\end{table}

\subsection{SMT Constraint Model Correctness}\label{sec:Constraint-Correctness}
We assert the correctness of the constraint model by presenting the following formal proofs.

\subsubsection{No Intraservice Remote Calls Constraint}
\begin{description}
    \item Goal: Prove $C_{nirc}$ ensures that a call from one endpoint to another can exist iff both endpoints are in different microservices.
    \item \textit{Axiom 1}. By definition, \texttt{E\_parent}($e_i$) gives the microservice that $e_i$ belongs to.
    \item \textit{Lemma 1}. By Axiom 1, $\texttt{E\_parent}(e_1)\ne \texttt{E\_parent}(e_2)$ asserts that $e_1$ and $e_2$ do not share the same parent microservices.
    \item \textit{Lemma 2}. By Lemma 1 and definition of $G$, $C_{nirc}$ asserts that $e_1$ and $e_2$ do not share the same parent microservice for each possible edge in $G$.
    \item \textit{Proposition 1}. By Lemma 2 and the fact that $G$ is a representation of $\mathcal{G}$, $C_{nirc}$ asserts that every call in $S$ (and extended versions of $S$) occurs from $\forall e_1 \in M_i$ to $\forall e_2 \in M_j$ where $M_i \ne M_j$.
    \item \textit{Theorem 1}. Based on Proposition 1, $C_{nirc}$ asserts that a call can only occur from an endpoint in one microservice to an endpoint in another unique microservice across all of $\mathcal{S}$ and its extensions.
\end{description}
\subsubsection{No Hub-like Dependencies Constraint}
\begin{description}
    \item Goal: Prove $C_{hub}$ ensures that no microservice has a sum of inbound and outbound calls to and from its endpoints that is greater than $\tau$.
    \item \textit{Corollary 1}. By definition of $conn_{E_a,E_b}$, $\sum_{E_j \in M_i, e \in E}(conn_{E_j,e} + conn_{e,E_j})$ sums all inbound connections to $E_j$ from any $e \in E$ and outbound connections from $E_j$ to any $e \in E$ for every $E_j \in M_i$.
    \item \textit{Lemma 3}. Based on Corollary 1 and definition of $\tau$, $C_{hub}$ asserts that the sum from Corollary 1 is less than $\tau$ for $\forall M_i \in M$.
    \item \textit{Theorem 2}. Based on Lemma 3, $C_{hub}$ ensures that no microservice in $\mathcal{S}$ and its extensions can have a sum of inbound and outbound calls to and from its endpoints that is greater than the threshold $\tau$.
\end{description}
\subsubsection{No Cyclic Dependencies Constraint}
\begin{description}
    \item Goal: Prove that $C_{cycle}$ ensures graph $G$ (and thus $\mathcal{G}$) contains no cycles.
    \item \textit{Axiom 2}. By definition of $L$ and $E$, every endpoint $E_i$ is assigned ordering $L_i$.
    \item \textit{Axiom 3}. $C_{cycle}$ asserts that, for each edge from $e$ to $e'$ in $G$, $L[e] < L[e']$.
    \item \textit{Corollary 4}. Assume there is a cycle present in $G$ consisting of a set of endpoints $Y$ and that $E_{initial}$ is some endpoint in the $Y$. Then $Y$ consists of endpoints such that there are edges from $E_{initial}$ to some $e_i \in Y$, which may have edges between other $e_j \in Y$ that eventually lead back to $E_{initial}$.
    \item \textit{Lemma 4}. Under the assumption of Corollary 4, applying Axioms 2 and 3 would mean that $L[e_{initial}] < L[e_i] < \ldots < L[e_j] < L[e_{initial}]$. Since $L[e_{initial}]<L[e_{initial}]$ is false for $\forall L[e_{initial}] \in \mathcal{Z}$ where $\mathcal{Z}$ is the set of integers, this is a contradiction.
    \item \textit{Proposition 2}. Based on Lemma 4, Axioms 2 and 3 hold iff there is no cycle present in $G$.
    \item \textit{Theorem 3}. Based on Proposition 2 and the definition of $C_{cycle}$, enforcing $C_{cycle}$ ensures $G$ has no cycles across the entirety of $\mathcal{S}$ and its extensions.
\end{description}

These proofs illustrate that the constraints enforce their respective rules for the model $\mathcal{S}$ and its extensions. Since we demonstrate the soundness of the model, we can also therefore assert that enforcing these rules for the model also enforces them on the actual microservice system encoded into $\mathcal{S}$ or one of its extended versions, formally illustrating the validity of both the intermediate model and constraint model.

\section{Research Questions}\label{sec:research-questions}
While we have outlined an extensible methodology for verifying concerns in microservice systems, a significant amount of future work is necessary to fully explore the potential and limitations of this methodology. We propose three research questions to guide this future work.

\textbf{RQ1:} How does this methodology's running time, compute resource usage, code coverage, and DORA metrics compare to existing methods that use static analysis (such as programmatically scanning systems and generating metrics) or dynamic analysis (such as performing end-to-end testing)?

\textbf{RQ2:} How does the methodology's runtime and compute resource usage scale as the number of microservices, number of endpoints, number of system connections, etc., increase? Can specialized solving, optimization, or incremental model construction techniques be applied to improve scalability? 

\textbf{RQ3:} What other concerns could our methodology be extended to address, and what constraints would need to be generated?

We propose RQ1 to quantify the potential benefit of our methodology. Meanwhile, we ask RQ2 to identify any potential limitations on tractability and efficiency as the number of concerns considered increases and the size of the microservice system grows larger. Finally, we identify RQ3 to increase the utility of the methodology by leveraging its extensibility.

These research questions can be addressed through a combination of formal proofs and empirical experiments on microservice system benchmarks or industry systems, when possible. Addressing these research questions will help extend the methodology and help remedy any current shortcomings of the methodology.

\section{Conclusion}\label{sec:Conclusion}
We advance an extensible methodology for software verification in microservice systems. While existing static analysis and runtime testing techniques can be used to verify concerns in microservice systems, our methodology has the potential to offer a more cost-effective, maintainable, and comprehensive approach to verification compared to these existing techniques.

Of course, this methodology is not meant to completely replace existing testing and static analysis methods in all use cases. For example, our methodology does not test functionality. However, it does verify that the system meets requirements for various concerns and has the potential to do this aspect of verification much more efficiently than testing or static analysis. While there are some potential limitations to our methodology, the novelty of applying formal methods for microservice system verification and the potential benefits over existing methods make this approach well worth exploring and evaluating.

\section*{Funding}\label{sec:Funding}
This work was supported by the National Science Foundation under Grant No. 2409933.

{
\bibliographystyle{IEEEtran}
\bibliography{IEEEabrv,main}

\begin{thebibliography}{10}
\providecommand{\url}[1]{#1}
\csname url@samestyle\endcsname
\providecommand{\newblock}{\relax}
\providecommand{\bibinfo}[2]{#2}
\providecommand{\BIBentrySTDinterwordspacing}{\spaceskip=0pt\relax}
\providecommand{\BIBentryALTinterwordstretchfactor}{4}
\providecommand{\BIBentryALTinterwordspacing}{\spaceskip=\fontdimen2\font plus
\BIBentryALTinterwordstretchfactor\fontdimen3\font minus \fontdimen4\font\relax}
\providecommand{\BIBforeignlanguage}[2]{{%
\expandafter\ifx\csname l@#1\endcsname\relax
\typeout{** WARNING: IEEEtran.bst: No hyphenation pattern has been}%
\typeout{** loaded for the language `#1'. Using the pattern for}%
\typeout{** the default language instead.}%
\else
\language=\csname l@#1\endcsname
\fi
#2}}
\providecommand{\BIBdecl}{\relax}
\BIBdecl

\bibitem{cerny2018microservice}
\BIBentryALTinterwordspacing
T.~Černý, M.~Donahoo, and M.~Trnka, ``Contextual understanding of microservice architecture: current and future directions,'' \emph{ACM SIGAPP Applied Computing Review}, vol.~17, pp. 29--45, Jan. 2018. [Online]. Available: \url{https://doi.org/10.1145/3183628.3183631}
\BIBentrySTDinterwordspacing

\bibitem{jawaddi2022formalReview}
\BIBentryALTinterwordspacing
S.~N.~A. Jawaddi, M.~H. Johari, and A.~Ismail, ``A review of microservices autoscaling with formal verification perspective,'' \emph{Software: Practice and Experience}, vol.~52, no.~11, pp. 2476--2495, 2022. [Online]. Available: \url{https://doi.org/10.1002/spe.3135}
\BIBentrySTDinterwordspacing

\bibitem{souri2018formal}
\BIBentryALTinterwordspacing
A.~Souri, N.~J. Navimipour, and A.~M. Rahmani, ``Formal verification approaches and standards in the cloud computing: a comprehensive and systematic review,'' \emph{Computer Standards \& Interfaces}, vol.~58, pp. 1--22, 2018. [Online]. Available: \url{https://doi.org/10.1016/j.csi.2017.11.007}
\BIBentrySTDinterwordspacing

\bibitem{amoroso2023msdev}
\BIBentryALTinterwordspacing
D.~Amoroso~d’Aragona, X.~Li, T.~Cerny, A.~Janes, V.~Lenarduzzi, and D.~Taibi, ``One microservice per developer: Is this the trend in {OSS}?'' in \emph{European Conference on Service-Oriented and Cloud Computing}.\hskip 1em plus 0.5em minus 0.4em\relax Springer Nature Switzerland, 2023, pp. 19--34. [Online]. Available: \url{https://doi.org/10.1007/978-3-031-46235-1_2}
\BIBentrySTDinterwordspacing

\bibitem{zhou2023microservicepain}
\BIBentryALTinterwordspacing
X.~Zhou, S.~Li, L.~Cao, H.~Zhang, Z.~Jia, C.~Zhong, Z.~Shan, and M.~A. Babar, ``Revisiting the practices and pains of microservice architecture in reality: An industrial inquiry,'' \emph{Journal of Systems and Software}, vol. 195, p. 111521, 2023. [Online]. Available: \url{https://doi.org/10.1016/j.jss.2022.111521}
\BIBentrySTDinterwordspacing

\bibitem{smith2023endtoendbenchmark}
\BIBentryALTinterwordspacing
S.~Smith, E.~Robinson, T.~Frederiksen, T.~Stevens, T.~Cerny, M.~Bures, and D.~Taibi, ``Benchmarks for end-to-end microservices testing,'' in \emph{2023 IEEE International Conference on Service-Oriented System Engineering (SOSE)}, 2023, pp. 60--66. [Online]. Available: \url{https://doi.org/10.1109/SOSE58276.2023.00013}
\BIBentrySTDinterwordspacing

\bibitem{cerny2025changeimpactanalysismicroservicesbased}
\BIBentryALTinterwordspacing
T.~Cerny, G.~Goulis, and A.~S. Abdelfattah, ``{Towards Change Impact Analysis in Microservices-based System Evolution},'' in \emph{2025 IEEE International Conference on Software Analysis, Evolution and Reengineering (SANER)}.\hskip 1em plus 0.5em minus 0.4em\relax Los Alamitos, CA, USA: IEEE Computer Society, Mar. 2025, pp. 159--169. [Online]. Available: \url{https://doi.org/10.1109/SANER64311.2025.00023}
\BIBentrySTDinterwordspacing

\bibitem{daraojimba2024DevOpsMetrics}
\BIBentryALTinterwordspacing
A.~I. Daraojimba, D.~Kisina, O.~S. Adanigbo, B.~C. Ubanadu, N.~A. Ochuba, and T.~P. Gbenle, ``Systematic review of key performance metrics in modern devops and software reliability engineering,'' \emph{International Journal of Future Engineering Innovations}, vol.~1, no.~1, pp. 101--107, 2024. [Online]. Available: \url{https://doi.org/10.54660/IJFEI.2024.1.1.101-107}
\BIBentrySTDinterwordspacing

\bibitem{Maruf2022TelemtryDynamic}
\BIBentryALTinterwordspacing
A.~Al~Maruf, A.~Bakhtin, T.~Cerny, and D.~Taibi, ``Using microservice telemetry data for system dynamic analysis,'' in \emph{2022 IEEE International Conference on Service-Oriented System Engineering (SOSE)}, 2022, pp. 29--38. [Online]. Available: \url{https://doi.org/10.1109/SOSE55356.2022.00010}
\BIBentrySTDinterwordspacing

\bibitem{Sotomayor2019RuntimeTesting}
\BIBentryALTinterwordspacing
J.~P. Sotomayor, S.~C. Allala, P.~Alt, J.~Phillips, T.~M. King, and P.~J. Clarke, ``Comparison of runtime testing tools for microservices,'' in \emph{2019 IEEE 43rd Annual Computer Software and Applications Conference (COMPSAC)}, vol.~2, 2019, pp. 356--361. [Online]. Available: \url{https://doi.org/10.1109/COMPSAC.2019.10232}
\BIBentrySTDinterwordspacing

\bibitem{cerny2023catalog}
\BIBentryALTinterwordspacing
T.~Cerny, A.~S. Abdelfattah, A.~A. Maruf, A.~Janes, and D.~Taibi, ``Catalog and detection techniques of microservice anti-patterns and bad smells: A tertiary study,'' \emph{Journal of Systems and Software}, vol. 206, p. 111829, 2023. [Online]. Available: \url{https://doi.org/10.1016/j.jss.2023.111829}
\BIBentrySTDinterwordspacing

\bibitem{abdelfattah2023towards}
\BIBentryALTinterwordspacing
A.~Abdelfattah, M.~Schiewe, J.~Curtis, T.~Cerny, and E.~Song, ``Towards security-aware microservices: On extracting endpoint data access operations to determine access rights,'' in \emph{Proceedings of the 13th International Conference on Cloud Computing and Services Science - CLOSER}, 2023, pp. 15--23. [Online]. Available: \url{https://doi.org/10.5220/0011707500003488}
\BIBentrySTDinterwordspacing

\bibitem{armando2010automatedsymbolicanalysisarbacpolicies}
\BIBentryALTinterwordspacing
A.~Armando and S.~Ranise, ``Automated symbolic analysis of arbac-policies (extended version),'' 2010, arXiv:1012.5590. [Online]. Available: \url{https://arxiv.org/abs/1012.5590}
\BIBentrySTDinterwordspacing

\bibitem{Lattuada2024FoundationForSysVerification}
\BIBentryALTinterwordspacing
A.~Lattuada, T.~Hance, J.~Bosamiya, M.~Brun, C.~Cho, H.~LeBlanc, P.~Srinivasan, R.~Achermann, T.~Chajed, C.~Hawblitzel, J.~Howell, J.~R. Lorch, O.~Padon, and B.~Parno, ``Verus: A practical foundation for systems verification,'' in \emph{Proceedings of the ACM SIGOPS 30th Symposium on Operating Systems Principles}, ser. SOSP '24.\hskip 1em plus 0.5em minus 0.4em\relax New York, NY, USA: Association for Computing Machinery, 2024, p. 438–454. [Online]. Available: \url{https://doi.org/10.1145/3694715.3695952}
\BIBentrySTDinterwordspacing

\bibitem{Backes2018FormalIAM}
\BIBentryALTinterwordspacing
J.~Backes, P.~Bolignano, B.~Cook, C.~Dodge, A.~Gacek, K.~Luckow, N.~Rungta, O.~Tkachuk, and C.~Varming, ``Semantic-based automated reasoning for {AWS} access policies using {SMT},'' in \emph{2018 Formal Methods in Computer Aided Design (FMCAD)}, 2018, pp. 1--9. [Online]. Available: \url{https://doi.org/10.23919/FMCAD.2018.8602994}
\BIBentrySTDinterwordspacing

\bibitem{panda2017SVinmicroservices}
\BIBentryALTinterwordspacing
A.~Panda, M.~Sagiv, and S.~Shenker, ``Verification in the age of microservices,'' in \emph{Proceedings of the 16th Workshop on Hot Topics in Operating Systems}, ser. HotOS '17.\hskip 1em plus 0.5em minus 0.4em\relax New York, NY, USA: Association for Computing Machinery, 2017, p. 30–36. [Online]. Available: \url{https://doi.org/10.1145/3102980.3102986}
\BIBentrySTDinterwordspacing

\bibitem{train-ticket-2018}
\BIBentryALTinterwordspacing
X.~Zhou, X.~Peng, T.~Xie, J.~Sun, C.~Xu, C.~Ji, and W.~Zhao, ``Benchmarking microservice systems for software engineering research,'' in \emph{Proceedings of the 40th International Conference on Software Engineering: Companion Proceeedings}, ser. ICSE '18.\hskip 1em plus 0.5em minus 0.4em\relax New York, NY, USA: Association for Computing Machinery, 2018, p. 323–324. [Online]. Available: \url{https://doi.org/10.1145/3183440.3194991}
\BIBentrySTDinterwordspacing

\bibitem{amoroso2024dataset}
\BIBentryALTinterwordspacing
D.~Amoroso~d'Aragona, A.~Bakhtin, X.~Li, R.~Su, L.~Adams, E.~Aponte, F.~Boyle, P.~Boyle, R.~Koerner, J.~Lee, F.~Tian, Y.~Wang, J.~Nyyss\"{o}l\"{a}, E.~Quevedo, S.~M. Rahaman, A.~S. Abdelfattah, M.~M\"{a}ntyl\"{a}, T.~Cerny, and D.~Taibi, ``A dataset of microservices-based open-source projects,'' in \emph{Proceedings of the 21st International Conference on Mining Software Repositories}, ser. MSR '24.\hskip 1em plus 0.5em minus 0.4em\relax New York, NY, USA: Association for Computing Machinery, 2024, p. 504–509. [Online]. Available: \url{https://doi.org/10.1145/3643991.3644890}
\BIBentrySTDinterwordspacing

\end{thebibliography}
}

\end{document}